# Monitoring of polymer viscosity by simultaneous ultrasonic and rheological measurements at high and varying temperatures

Nesrine Houhat[1*], Thibaut Devaux[1], Samuel Callé[1], Laksana Saengdee[1], Séverine Boucaud Gauchet[1], François Vander Meulen[1]

[1]GREMAN UMR 7347, INSA Centre Val de la Loire, Université de Tours, CNRS, 41000 Blois, France.
*Corresponding author: nesrine.houhat@insa-cvl.fr

**Abstract**

**A thorough comprehension of the rheological behavior of polymers during industrial processes is essential for optimizing manufacturing efficiency and product quality. The final properties and behavior of resulting polymer parts are known to be directly linked to the thermomechanical evolution of materials during their processing. The non-invasive monitoring of this stage could improve the quality of manufactured equipment. This can be done by tracking viscosity off-line on a rheometer. In this article, an experimental method to monitor the viscosity of polymer material at high and varying temperatures by using ultrasound is proposed. This method allows to measure ultrasonic and rheological properties of a sample, simultaneously and in real-time. An ultrasonic instrumentation adapted to a rheometer for continuous monitoring. It allows high-temperature range measurements (up to 200°C). A dedicated signal processing algorithm is developed to determine the polymer longitudinal acoustic velocity by considering wave packets overlapping and temperature variation. Results on polyethylene show that ultrasonic parameters appear to be sensitive to changes in polymer state. It enables more accurate detection of the onset of polymer crystallization. The study paves the way for ultrasonic real-time monitoring of the rotomolding process.**

## 1. Introduction

Rheology plays a fundamental role in the processing of polymer melts. Understanding the rheological properties of polymers enables to optimize transformation processes and to ensure the quality of finished products as well as to improve the efficiency of industrial tools [1-2]. Polymers become viscous and malleable when heated above their melting temperature. The way they flow under the effect of applied forces (thermal, mechanical) is a key factor in transformation processes. Process parameters allow to monitor the process stability in terms of flow and heat transfer conditions, but not directly to the material quality. Flow simulations using multiphysics models are used to optimize the process and to predict the rheological behavior of polymer melts as a function of processing conditions [3], thus determining the optimum processability window to reduce development costs. Currently, the rheological properties of polymers and their behavior are experimentally characterized off-line of the process, using rheometers or viscometers. This indicates a significant difference between production and detection of material quality problems. The rheological behavior of polymers depends on processing behavior and it may be influenced by sample preparation for off-line measurements. The experimental validation of established simulation models is biased due to the off-line situation. For these reasons, in-line and on-line measurements of the rheological properties of polymers and compounds during processing has become an important objective for both scientific work and industrial applications for decades [4-7]. Several studies of in and on-line measurements during extrusion and also injection molding can be found in the literature [8-11], with on-line systems where a part of the melt is diverting away from the process to e.g., a rheometer, and in-line systems for which the estimation or measurement is conducted in the actual injection molding process.

Among the main polymer transformation processes, rotational molding is chosen for this study [12]. It is an alternative to blow molding and the injection welding for manufacturing polyethylene liner of high-pressure hydrogen storage tanks [12-15]. Rotational molding is a process used to produce small or large hollow parts with complex shapes from thermoplastic polymers. In this process, a powdered polymer in a mold is rotated on two perpendicular axes at low speeds, heated until the polymer melts and coats the mold's inner surface, then cooled to solidify [16-18]. The polymer then undergoes two changes of state [19-20]: melting (with sintering and densification) and solidification (with crystallization and shrinkage) [21-25]. Unlike for another molding processes like extrusion or injection, no device is currently described in the literature for in-line measurement of polymer viscosity during the rotational molding process. The only parameter currently measured and controlled during this process is the temperature by using tools like Rotolog® and Datapack® [16-26]. It allows indirect monitoring of the polymer physical state by measuring the mold internal air temperature over time.

Optimizing the polymer transformation process requires the direct non-invasive measurement of the viscoelastic properties at each step of the process. The in-line monitoring in real time and without contact of rheological properties requires the implementation of an innovative acquisition system to probe the material and adapted to the constraint of temperature range of polymer processing. For this device to be viable, it is essential that the measurements do not interfere with the process and vice versa. Ultrasonic nondestructive methods allow the direct measurement of material elastic properties. The advantages of ultrasonic nondestructive techniques lie in their relative cost-effectiveness, high sensitivity, and their ability to perform in-line and real-time measurements. The use of these techniques finds extensive application for the detection of dynamic transformations and transition points in polymer materials [27-29]. Meanwhile, the reliability of studying the curing process at room temperature of thermosetting resins using ultrasounds methods has been demonstrated [30-31]. During this curing, the viscosity of polymer changes strongly: the viscosity of polymer decreases with the elevation of temperature. When the curing reaction is sufficiently advanced, the viscosity begins to increase until the polymer solidifies completely. Experimental results have shown an evolution on velocity and attenuation during the curing process [30-31].

As well, many studies based on the acoustic methods have been conducted to investigate the quality of rotomolded solid parts at the rotomolding machine output. The defects sought and quantified are porosities, formed during the sintering phase, and thickness variations [32-35]. Measuring the melt viscosity in the polymer rotational molding process with ultrasonic device provides knowledge to prevent defect formation during rotomolding manufacture. Ultrasounds have already demonstrated their robustness for tracking and monitoring polymer manufacturing process [35]. All these studies demonstrate that ultrasound methods seem to be an adapted technology to monitor the evolution of the rheological properties during rotational molding cycle. Ultrasonic sensors have already been used to measure the real temperature of melt polymer during injection process [36-38]. To our knowledge, the literature is lacking on the use of ultrasonic instrumentation to describe the evolution of the viscoelastic properties of a polymer during the rotational molding process.

To address this gap in the literature, the understanding of the polymer properties during the manufacturing process transformation should be studied. As rheology is a key factor in the processing of polymer melts, we propose, in this paper an *ex-situ* experimental study of the relationship between the rheological properties and the ultrasonic properties of the melt polymer at high and varying temperatures. The objective of the present study is to develop a specific acoustic instrumentation within an oscillatory rheometer chamber. To mimic the varying temperature behavior during rotational molding process of thermoplastic polymer, the acoustic instrumentation is adapted for high temperature monitoring (above 200°C).

The two technical and scientific challenges are: *i)* the achievement of ultrasonic measurement in a single access space requiring a small footprint, and *ii)* the identification of relationships between the parameters provided by the rheometer and those provided by the ultrasonic measurements associated to a dedicated signal processing. This allows to simultaneously perform ultrasonic and rheological measurements. The experimental setup is first validated on silicone oils with a calibrated viscosity in a temperature range varying from 25°C to 85°C. In a second time, measurements have been done at a high temperature condition on polyethylene.

Following the introduction, the manuscript is then organized in three sections. The first section presents the materials, the rheological tests and the developed ultrasonic instrumentation on the oscillatory rheometer. The second section is devoted to describe the signal post-processing method specifically adapted to measure ultrasonic longitudinal velocity within the thin polymer layer and considering the temperature changes. The last section is dedicated to the experimental results on silicone oils and polyethylene. Correlation of ultrasonic measurements with rheological parameters and the discussion of the results are then presented.

## 2. Rheological and ultrasonic methods

This section presents the studied materials, and the rheological tests used for the experimental characterization. The developed ultrasonic instrumentation of the rheometer is also described.

### 2.1 Materials

To validate the experimental apparatus and the associated post-processing algorithm, calibrated silicone oils are first used as testing materials allowing to perform experiments from 25°C to 85°C without state change.

Silicone oils (polydimethylsiloxane-PDMS) with four different viscosities are used: 1 Pa.s, 10 Pa.s, 100 Pa.s and 1000 Pa.s at 25°C. They are provided by Sigma-Aldrich except the 1000 Pa.s silicone oil which is supplied by Wacker.

The medium density polyethylene (MDPE) is studied during its change of state in temperature range reaching 200°C. Polyethylene Revolve 6402U employed in this research is a powder supplied by Arkema with a density of 0.932g/cm$^3$ and a melt flow index of 4g/10min. Melt temperature and crystallization temperature have been measured by Differential Scanning Calorimetry (DSC) at 125°C and 115°C respectively with a heating and cooling rate of 5°C/min.

## 2.2 Rheological characterization method

The rheological measurements have been carried out as a controlled-shear-strain test on an oscillatory Anton Paar Rheometer (MCR302). Molten polymers have a rheological behavior known as viscoelastic. They are characterized by both viscous flow and elastic deformation. In the case of molten polymers, determining viscosity is not enough, as elastic effects can occur during polymer transformation. Rheological measurements are carried out in oscillatory mode with a sinusoidal curve. This loading mode allows both the elastic part of the polymer's viscoelastic behavior to be considered by measuring the storage modulus G' and the viscous part of the viscoelastic behavior by measuring the loss modulus G". The law of elasticity described by equation 1 is applied to oscillatory shear tests with the complex shear modulus G* describing the complete viscoelastic behavior of the polymer. The loss or damping factor tan δ is defined by the ratio between the two parts of the viscoelastic behavior (equation 2). δ corresponds to the phase shift angle between the sinusoidal input excitation signal and the sinusoidal output signal. Remember that ideal elastic behavior corresponds to a value of δ=0 and ideal viscous behavior to a value of δ=90°. In the case of molten polymers, δ is comprised between 0 and 90°.

$$G^* = \frac{\tau_A}{\gamma_A} \qquad (1)$$

with the shear stress amplitude $\tau_A$ (in Pa) and shear strain amplitude $\gamma_A$ (dimensionless)

$$\tan \delta = \frac{G''}{G'} \qquad (2)$$

The complex viscosity |η*| is defined by equation 2.3. In the context of the Maxwellian behavior of polymer melts where G">G', the complex viscosity remains constant at low frequencies. The viscosity |η*| is then equal to the zero-shear viscosity $\eta_0$ measured by rotation.

$$|\eta^*| = \frac{G^*}{\omega} \qquad (3)$$

with complex shear modulus G* (in Pa) and angular frequency ω (in $s^{-1}$).

The complex viscosity (|η*|), the temperature T and the gap width *e* data are collected by the rheological software (RheoCompass) on a dedicated computer. To minimize the temperature difference within the measuring chamber which contains the sample, an active hood where the temperature is controlled is used in addition to Peltier elements. This and the small amount of polymer of the sample ensure that, given the recommendation of Anton Paar, the temperature in the sample can be considered homogeneous during the process.

For analyzing the temperature-dependent of the complex viscosity |η*|, the shearing in oscillatory tests is performed with plate/plate measuring system under constant dynamic-mechanical conditions. Both amplitude of deformation and frequency are kept constant, and the temperature profile is preset with a linear temperature increase or decrease over the duration of the measuring interval. The circular rotating plate is 25 mm diameter. The gap width *e* varies

from 0.5 to 1.2mm as recommended by the standards ISO6721-10 and DIN 53019. This gap width corresponds to the thickness of studied polymer sample $e$. The parallel plate geometry is adapted to characterize the rheological properties of high viscosity polymer melts. It offers greater flexibility in adjusting the gap, accommodating thermal expansion and shrinkage. Moreover, it is suitable for ultrasonic measurements where a flat, parallel surface is required for optimal signal transmission.

The measuring system is composed of a quartz inferior plate and a metallic upper. Initially, this system has been developed by Anton Paar to study the sample flow by optical microscopy which is positioned below the lower plate during the shearing test with a controlled temperature.

The silicone oils present a Newtonian flow behavior while molt MDPE has a shear-thinning flow behavior. The conditions of the temperature sweep are established from some preliminary tests. The complex viscosity $|\eta^*|$ is equivalent to the zero-shear viscosity $\eta_0$ at the amplitude deformation value selected within the Linear ViscoElastic (LVE) region and in the lower frequency range. In these conditions, the viscosity is constant and independent of the amplitude deformation and the frequency range. The temperature sweep conditions to study silicone oil temperature-dependent behavior in the range of 25-85°C are a constant gap width $e$ fixed to 1mm, a controlled-shear-strain of 1%, an angular frequency $\omega$ of 1 rad/s and a heating rate of 5°C/min. The measurement duration is 12 minutes.

The temperature-dependent behavior of MDPE is used for investigating melting behavior of MDPE when heated, and the crystallization when cooled. The temperature profile during temperature sweep is defined by a temperature gradient of 5°C/min. Within the first step, polymer is heated from 160°C to 200°C. During the second step, the polymer is cooled down from 200°C to 115°C. A 5-minute temperature hold is observed between the two steps. A rheological measurement cannot be achieved under 115°C because of the MDPE solidification. The gap width $e$ is not fixed and it varies during the test to consider the dimensional variation induced by the dilatation and the shrinkage of polymer sample during the temperature variations. The initial gap width is 1.2mm. The shear strain is 0.1% and the angular frequency $\omega$ is 1 rad/s and the temperature rate is 5°C/min.

Before the heating temperature-dependent behavior of MDPE, the polymer is heated to 200°C and sheared at this temperature to annihilate bubbles forming within the molt sample. Afterward, the temperature is reduced to 160°C by 5°C/min, followed by a 5-minute temperature equilibration period. Simultaneous rheological and ultrasonic measurements start at this moment. The MDPE sample change of state is happening fast and a delay of 6s between two measurements has been found to be enough to capture the evolution of the sample. The total measurement time is about 60 min including preparation, heating and cooling regimes.

### 2.3 Ultrasonic instrumentation

Simultaneously with the rheological tests described previously, ultrasonic measurements have been carried out. The ultrasonic experimental setup used to characterize the polymer samples on rheometer is depicted in Figure 1.

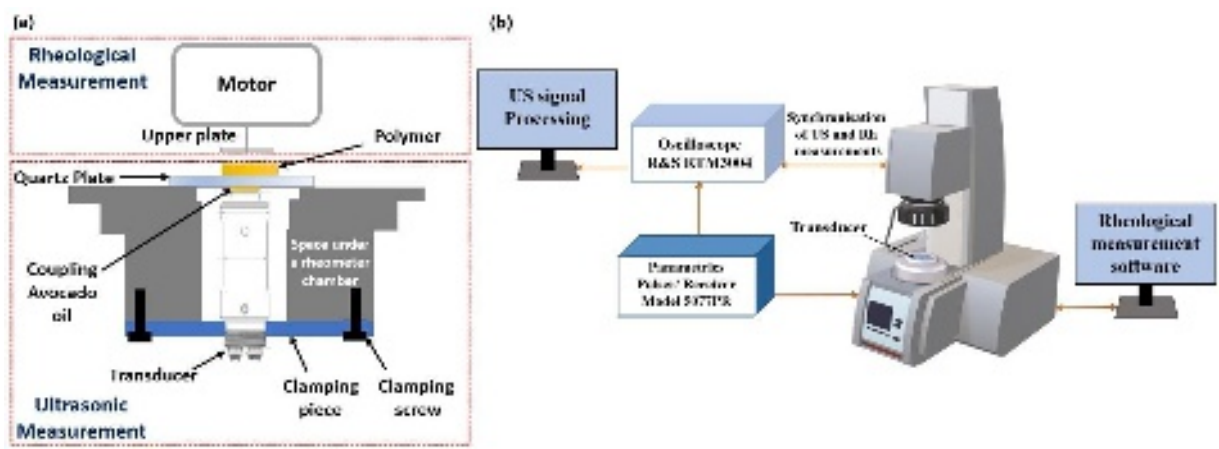

**Figure 1.** Scheme of the ultrasonic instrumentation of the rheometer. (a) To simultaneously measure the viscosity and the ultrasonic properties of studied sample, an acoustic apparatus composed of a transducer and a clamping piece are added to the rheometer. The avocado oil ensures the acoustic transmission at high temperature (up to 200°C) between transducer and the quartz plate where the studied sample is set. (b) Rheological measurements are collected using a dedicated software (RheoCompass). Ultrasound measurements are performed in pulse-echo method and synchronized with rheological measurements.

A transducer is placed in a space under the rheometer chamber supported by a metallic plate designed to adapt to the rheometer geometry (Figure 1(a)). It is fixed with two screws inserted in tapped holes already present in the rheometer. The metallic plate includes a centered hole to support the transducer. The transducer is in contact under a 4mm thick quartz plate supporting the tested material.

As measurements with polymers involve high temperatures (up to 200°C), a high temperature resisting transducer is used (Ionix HS582i). It allows measurements for a wide measurement temperature range from -55 to +550°C with 5MHz central frequency. To ensure a stable acoustic transmission at the quartz / transducer interface during high temperature measurement, a specific coupling material is used. During experiments, avocado oil has been chosen as coupling material due to its high boiling temperature ~270°C. The acoustic properties are assumed to be constant during all measurement time.

The electrical impulse is generated by a Squarewave Pulser/Receiver (Panametrics, Model 5077PR) as shown in Figure 1(b). The signal is received by the transducer in echo mode and then digitized by an oscilloscope (Rohde and Schwarz RTM3004). The digitized signal is transmitted to a computer in real time for signal processing and storage. Rheometer output channel is connected to the oscilloscope to synchronize ultrasonic and rheological measurements.

The measured temperature on the rheometer corresponds to a measurement number $j$ during heating and cooling steps. The temperature values are measured by the rheometer and correspond to the temperature in the chamber. The sample is considered small enough to neglect the difference between its internal temperature and the chamber. For each rheological measurement $j$, the rheometer generates a trigger signal which is detected by the oscilloscope. A dedicated Matlab program allows the synchronization between rheological and ultrasonic measurement. One ultrasonic measurement is achieved for each rheological measurement $j$, with an average of 256 successive acquisitions.

## 3. Ultrasonic signal processing

### 3.1 Time of flight measurement method

To measure the polymer echo TOF in the rheometer, two configurations have been considered as shown on figure 2. The first one corresponds to the reference, with no polymer in the rheometer. The received signal is given by:

$$S_q(t) = L_q(t) + \sum_n Q_{q,n}(t), \tag{4}$$

where the subscript $q$ denotes the case with quartz only, $L_q(t)$ is the echo reflected at the interface transducer delay line/quartz plate and $Q_{q,n}(t)$ are the $n$ back and forth echoes reflected at the interface quartz/air.

The second configuration corresponds to the case where the tested polymer is on the quartz plate. The received signal noted $S_{q+p}(t)$ is:

$$S_{q+p}(t) = L_{q+p}(t) + \sum_n Q_{q+p,n}(t) + E_1(t), \tag{5}$$

where subscript $q+p$ denotes the case with quartz and polymer. $E_1(t)$ is the first echo propagating in the polymer layer and reflected by the upper metallic plate.

The ultrasonic velocity is estimated from the TOF and the thickness of the polymer which is obtained from the gap measurement.

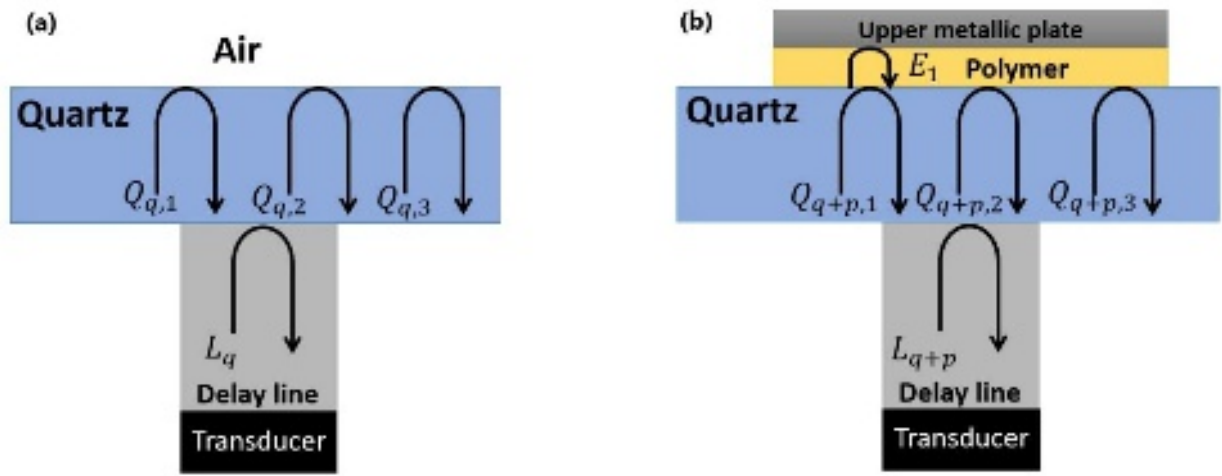

Figure 2. Ultrasonic velocity determination principle: (a) reference measurement without polymer, (b) measurement with the polymer layer in the rheometer. Arrows in the figures indicates the acoustic paths.

Given the low polymer thickness (~1mm) involved in the rheometer, the echo $E_1(t)$ is delayed of about 1μs from the quartz signal $Q_{q+p,2}(t)$, as shown in Figure 3(a). This leads to an overlapping of these two echoes, making detection and analysis of $E_1(t)$ difficult [39-45]. To overcome this problem, the reference signal $S_q(t)$ is subtracted from the signal propagating in the polymer $S_{q+p}(t)$. Then, it allows us to identify wave packets and to calculate the TOF to determine the acoustic velocity. An example of the result in the 10 Pa.s silicone oil at 25°C is shown on figure 3(a).

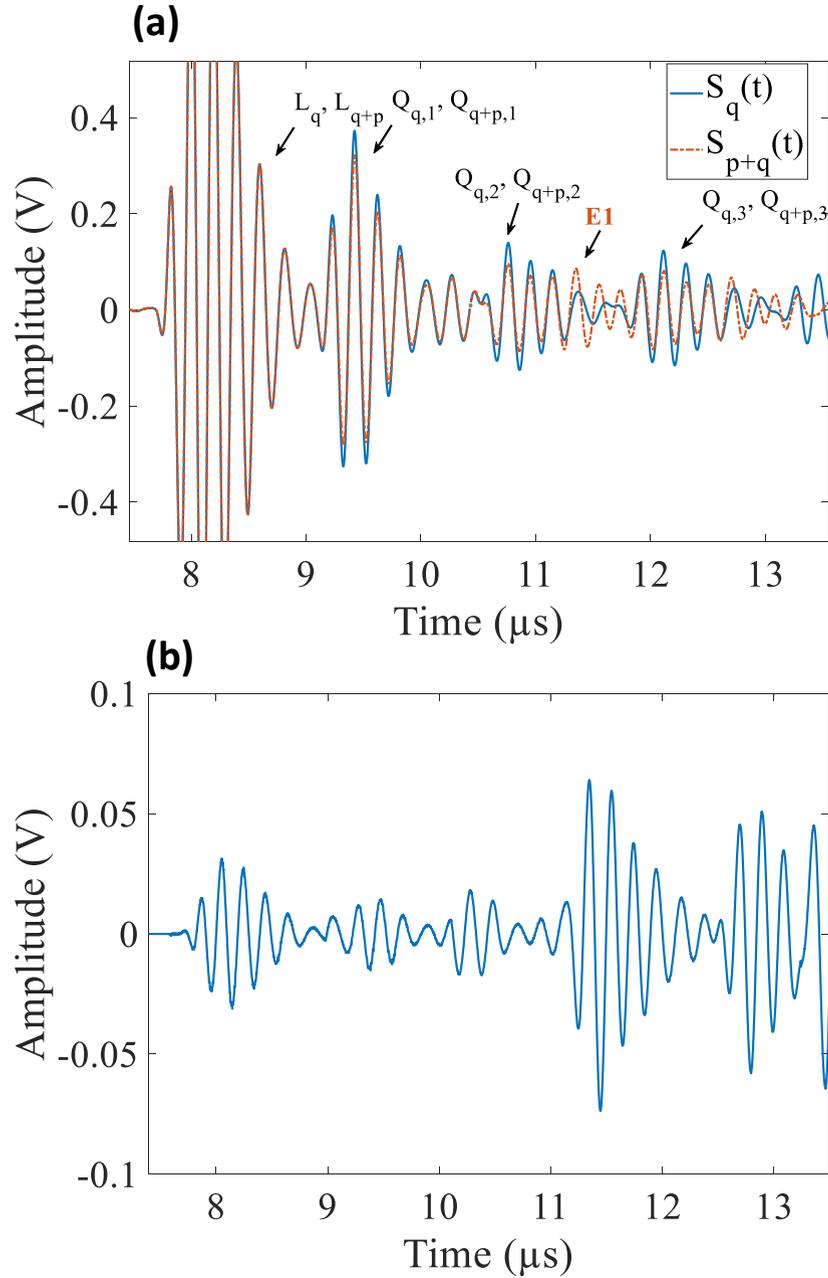

Figure 3. (a) Temporal signal received from quartz $S_q(t)$ (blue line) and with the 10 Pa.s silicone oil layer $S_{q+p}(t)$ (orange dashed line) at 25°C. The signal $E_1(t)$ corresponding to the first back and forth in the silicone oil overlaps with the quartz echo $Q_{q+p,2}$, (b) the corresponding calculated difference signal $S_{diff}(t)$.

However, the reference signal $L_q(t) + Q_q(t)$ differs from signal $L_{q+p}(t) + Q_{q+p}(t)$. During the rheological measurements, the temperature varies, and as a consequence, the ultrasonic velocities in the quartz plate varies too. The transmission coefficients in and out the quartz plate are thus affected too. These variations induce signal time shifts, amplitude and frequency variations. Moreover, the interface change from quartz-air to a quartz-polymer gives rise to amplitude discrepancies between $Q_q(t)$ and $Q_{q+p}(t)$. To tackle this limitation, a signal processing procedure to shift in time and scale the acquired signals has been developed. This method is presented in details in the following section.

## 3.2 Procedure description for E₁(t) determination

To evaluate the ultrasonic velocity in the polymer, a determination of the time of arrival of E₁(t) echo, which is overlapped with $Q_{q+p}(t)$ is necessary. Figure 4 shows the flowchart of the adopted processing method for the determination of the ultrasonic TOF in the polymer thickness.

This method is based on the subtraction of the reference signal received without the polymer layer from the scaled and time shifted signal obtained with the polymer. The time shifting has been achieved by intercorrelation and the amplitude readjustment has been done by calculating a weighting amplitude coefficient corresponding to each wave packet. In the following, the different steps of the proposed method are presented.

### *Rheological and ultrasonic measurements*

First, the ultrasonic reference signal $S_q(t)$ is measured at a fixed temperature and windowed by a rectangular window according to the different signal wave packets noted $S_q^i(t)$, where $i = 1,2,...,n$ denotes the number of the considered wave packet. In our case, the time needed to go through the quartz plate is around 9µs and the time needed for the wave to propagate back and forth in the polymer layer is around 11µs. The complete temporal length of the signal we have to process is 13.5µs (see figure 3(a)). n=4 is enough to acquire the relevant signal E₁(t). The rheological and ultrasonic measurements are performed at fixed time increments $t_j$, with $1 \geq j \geq j_{max}$. Simultaneous measurements of viscosity $\eta^j$, temperature $T^j$, gap $e_p^j$ and signals $S_{q+p}^j(t)$ are done. The following procedure is adopted for each time increment $j$.

### *Time shifting*

As mentioned in section 3.1, signals acquired during temperature sweeps are time shifted compared to the reference signal. To remove the temperature effect, the signal $S_{q+p}^j(t)$ is windowed (by a rectangular window) to extract the echo $S_{q+p}^{1,j}(t)$ corresponding to the reflection $L'_{q+p}(t)$ on the transducer delay line/quartz interface. This echo is also extracted from the $S_q(t)$ signal, corresponding to $S_q^1(t)$. The cross-correlation between these two signals allows to determine a time offset $\tau_j$, which will be used to shift $S_q^i(t)$ and $S_{q+p}^j(t)$. Same windows are applied to the shifted signal $S_{q+p}^j(t - \tau_j)$ leading to $S_{q+p}^{i,j}(t)$.

### *Amplitude scaling*

As the transmission coefficients through the quartz are varying because of temperature and interface changes, the signals $S_{q+p}^j(t - \tau_j)$ have different amplitudes according to the time increment of acquisition $j$. To suppress quartz echoes to better distinguish the polymer echo E₁(t), the signal is amplitude-adjusted by multiplying each window by the amplitude weighting coefficient $\alpha_i^j$ equal to the ratio of the quartz echo maxima for the $S_q^i(t)$ and $S_{q+p}^{i,j}(t)$ acquisitions. Thus, the variation of transmission coefficient in quartz is considered in the suppression of the quartz echoes.

*Calculation of the difference signal*

The difference signal $S_{diff}^{j}(t)$ is obtained by subtracting the reference signal from the time-shifted and amplitude-scaled signal. It is given by the expression:

$$S_{diff}^{j}(t) = \sum_{i=0}^{n}(\alpha_i^j S_{q+p}^{i,j}(t-\tau_j) - S_q^i(t)) \qquad (6)$$

The difference signal thus obtained lead to the estimation of E₁(t). The quartz echoes $Q_{p+q,n}(t)$ have been subtracted.

*TOF in polymer*

The envelopes of the difference signal $S_{diff}^{j}(t)$ and the signal $S_q^2(t)$ are calculated. The times of the envelope maxima ( $t_{ref}$ and $t_{diff}^{j}$) are determined using the second order interpolation function to enhanced the time resolution. From the time of flight in the polymer $t_P^j = t_{diff}^{j} - t_{ref}$, the ultrasonic velocity is estimated:

$$V_p^j = \frac{2e_p^j}{t_P^j} \qquad (7)$$

The procedure is repeated until $j_{max}$. The end of the rheological measurements is due to the solidification of the polymer in the case of MDPE.

The following section gives the ultrasonic velocity results obtained by adopting this procedure as a function of temperature for the silicone oil and the MDPE.

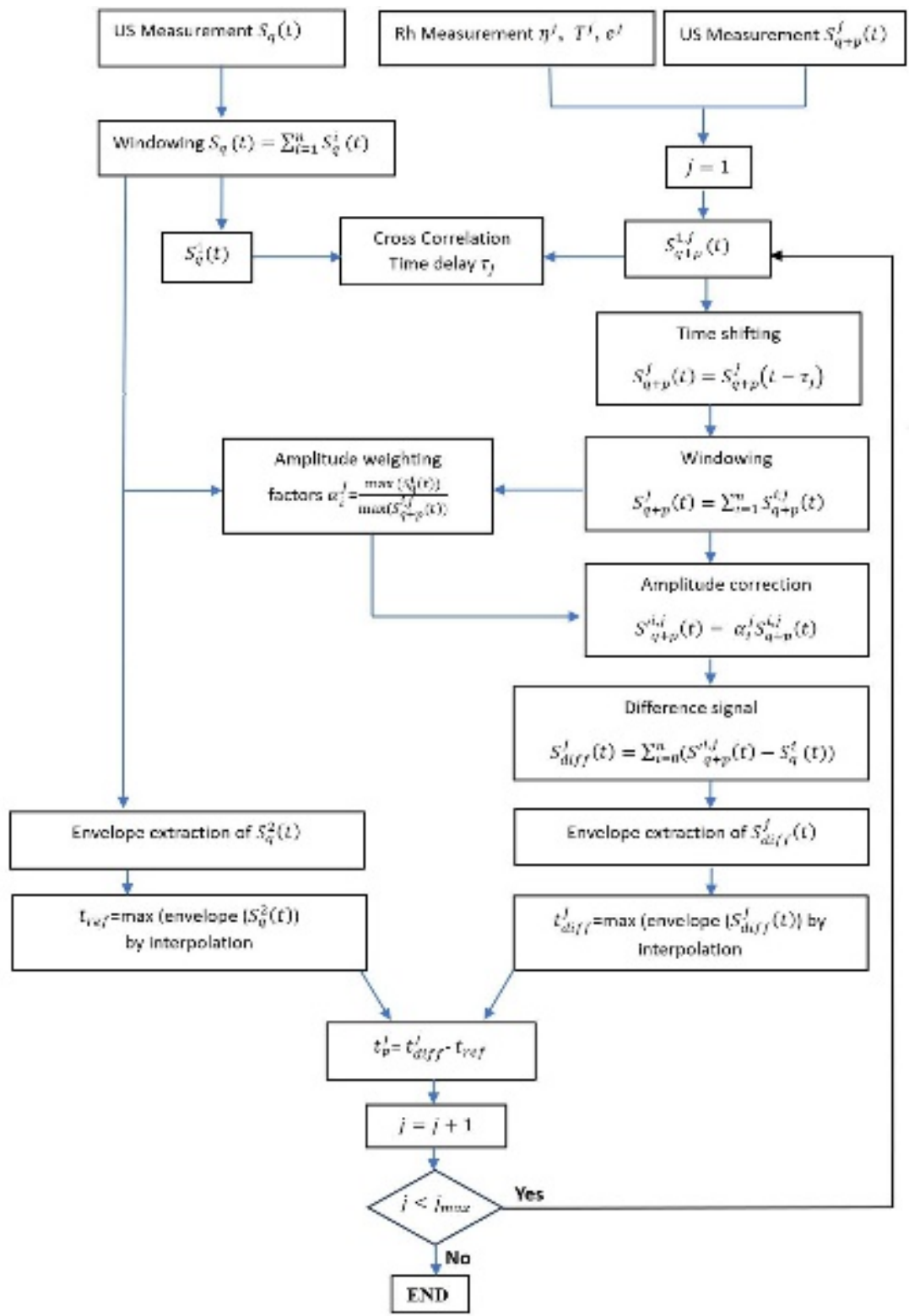

Figure 4. Flowchart of the signal processing method for the determination of the time of flight (TOF) of the first echo propagating in the polymer layer E1(t). This method is based on the subtraction of the ultrasonic reference signal $S_q(t)$ without the polymer layer from the signal obtained with polymer layer at different temperatures $T_j$ with subsequent processing for quartz wave packets $S_q^i(t)$. The difference signal $S_{diff}^j(t)$ is then calculated by subtracting the reference signal from the time-shifted and amplitude-scaled polymer signal to account for temperature-induced variations in transmission coefficients through quartz. Time of flight (TOF) in the polymer is determined by analyzing the envelopes of $S_{diff}^j(t)$ and $S_q^2(t)$ corresponding to the first quartz echo following calculation of ultrasonic velocity $V_p^j$. This procedure is repeated until $j_{max}$, marking the end of rheological measurements due to polymer solidification.

## 4. Results and discussion

The proposed signal processing method is implemented on the described experimental setup in section 2. The newly designed ultrasonic apparatus and signal processing method have first been validated in liquid polymer (silicone oils) between 25°C and 85°C and then used during the study of heating and cooling dependent behavior of MDPE with a maximal temperature reaching 200°C.

### 4.1. Validation with silicone oils

#### 4.1.1 Complex viscosity measurements

Figure 5 shows the measured complex viscosity $|\eta^*|$ during a temperature rise from 25 to 85 °C for the 1, 10, 100 Pa.s silicone oils and from 30 to 85°C for the 1000 Pa.s silicone oil. In this temperature range, viscosities vary due to temperature change while their physical state remains unchanged [46]. It can be seen that the complex viscosity measurements at 25°C for the 1, 10 and 100 Pa.s silicone oils are in good agreement with the nominal values. Moreover, the viscosity values are decreasing with the increase of temperature [47]. Discrepancy for the rheological measurements of the 1 Pa.s silicone oil could be explained by the lack of precision due the low viscosity. Using a plate disk measuring system with a bigger diameter 50 mm should improve the measurement.

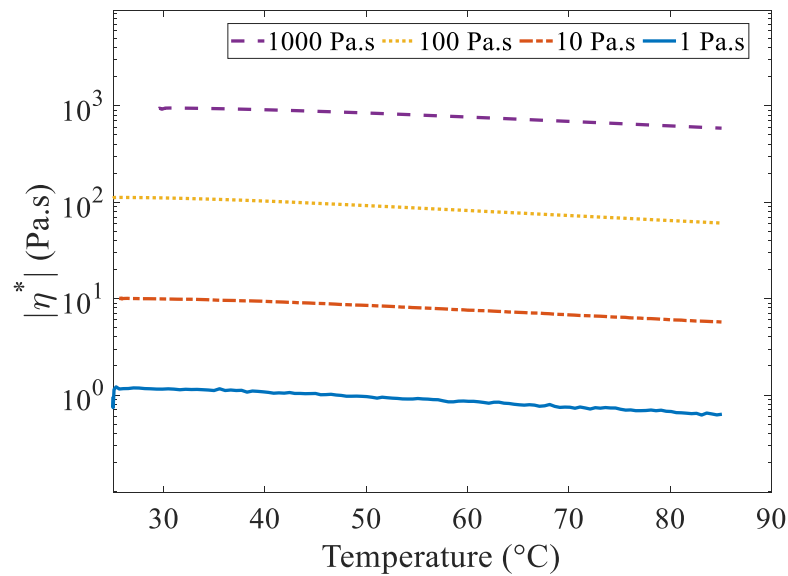

Figure 5. Complex viscosity $|\eta^*|$ as a function of temperature from 25°C to 85°C for the 1, 10, 100 and 1000 Pa.s silicone oils.

#### 4.1.2 Ultrasonic velocity measurements

Figures 6(a) and 6(b) illustrate respectively the obtained signal from the 10 Pa.s silicone oil at 85°C and the corresponding difference signal obtained after the signal post-processing described in section 3. In Figure 6(a), the first echo (before 9.5 µs) corresponds to the first reflection at the interface quartz/silicone oil. From 10.5 µs, the signal $S_{q+p}(t)$ is an overlapping of the back-and-forth echoes in the quartz plate and those propagating in the silicone oil. To distinguish the silicone oil echoes, the difference signal (figure 6(b)) is computed following the signal processing method described in section 3.2. The TOF is computed between this signal and the first echo propagating in the quartz plate.

Figures 6(c) and 6(d) show respectively a cartography of the envelope of the 10 Pa.s silicone oil signals and the envelope of the difference signal obtained from 25°C to 85°C. The red line corresponds to the time of the maximum envelope of $E_1(t)$. It can be noted that the amplitude of the quartz echoes is reduced in the difference signals compared to their amplitude in the signals $S_{p+q}(t)$. This makes it easier to distinguish $E_1(t)$ from the difference signal. Moreover, the temperature influence on the arrival time of $E_1(t)$ which increases with temperature is easily observable. The echo arriving at t=13µs at 85°C corresponds to one back and forth in the polymer and one back and forth in the quartz plate.

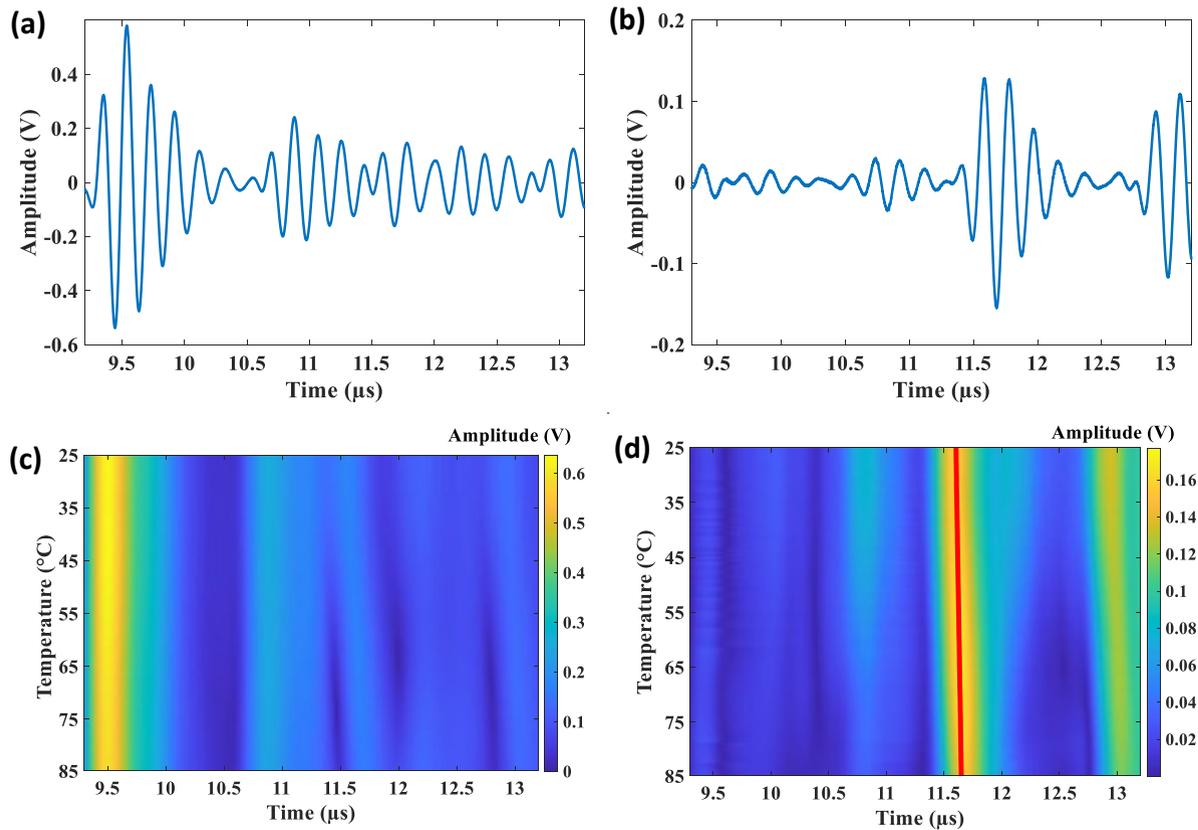

Figure 6. (a) Temporal signal obtained from the 10 Pa.s silicone oil at 85°C, (b) The corresponding difference signal calculated by the method described in section 3.2, (c) Cartography of the envelope of the time shifted signals obtained from 25°C to 85°C, (d) Cartography of the envelope of the corresponding difference signals. The red line represents the time of the maximum envelope of $E_1(t)$ echo.

Figure 7 shows ultrasonic (US) velocity variations during a temperature rise from 25 to 85 °C for the 1,10,100 and 1000 Pa.s silicone oils. It is derived from the TOF estimation (red line in Figure 6(d)) and the measured gap in the rheometer.

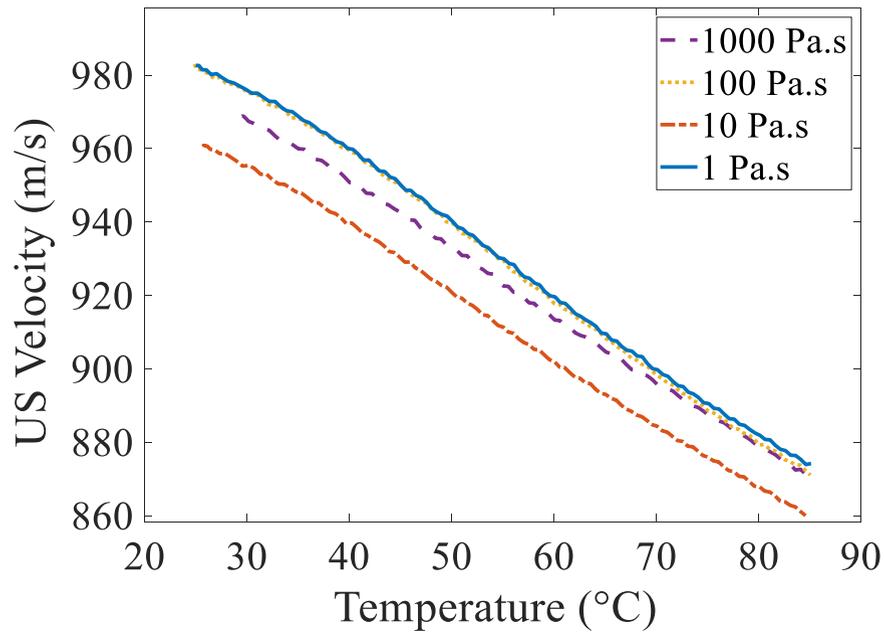

Figure 7. Ultrasonic velocity versus temperature from 25°C to 85°C for the 1,10,100 and 1000 Pa.s silicone oils.

Figure 7 depicts US velocities decrease of about 1.6 m. s$^{-1}$/°C for each viscosity as expected [48]. The velocity values obtained for the four silicone oils at 25°C are used as a reference for the evaluation of the relative ultrasonic velocity as a function of temperature. Similarly, the complex viscosities measured and reported in Table.1 at 25°C are used for the relative complex velocity evaluation at the upper temperatures.

Independently, velocity measurements are performed at 25°C in transmission by a conventional insertion-substitution method [49] to validate the *in-situ* measurements. Two paired transducers of 2.25MHz were used for transmitting and receiving ultrasonic waves in a water tank. This first measurement provides a reference measurement which allows to estimate the wave path distance between the two transducers since the ultrasonic velocity is known. Then, the water is substituted by the silicon oils and the ultrasonic velocity is estimated knowing the TOF and the distance calculated previously.

Table 1 shows the ultrasonic velocity results obtained at 5MHz for the four silicone oils obtained by the described method in section 3 and those obtained by a conventional insertion substitution method. The density and viscosity measurements have been performed and reported in the Table 1. Measurements have not been performed on the 1000 Pa.s silicone oil because of its high viscosity. The difference of density for the three silicone oils is not significant. The relative velocity difference between the values obtained from the insertion substitution method and those measured *in-situ* in the rheometer is presented. The complex viscosity is calculated from to the measured density and kinematic viscosity of oil following the relation:

$$\nu = \frac{\eta_0}{\rho} \qquad (8)$$

with ν, the kinematic viscosity (m²/s), $\eta_0$, the zero-shear-viscosity (Pa.s) and ρ, the density (kg/m³). In our test conditions, the complex viscosity and zero-shear-viscosity have the same value.

Table.1: Measured US velocities, densities and viscosities of the various silicone oils used at 25°C

| Nominal Silicone oil viscosity (Pa.s) | 1 | 10 | 100 | 1000 |
|---|---|---|---|---|
| Measured complex viscosity (Pa.s) | 1.14 | 10.12 | 111.7 | / |
| Measured Density (kg/m³) | 927 | 949 | 966 | / |
| Calculated complex viscosity (Pa.s) | 0.93 | 9.49 | 96.6 | / |
| Ultrasonic velocity (m/s) by insertion substitution | 1017 | 1024 | 1016 | / |
| Ultrasonic velocity (m/s) measured *in-situ* | 983 | 960 | 981 | 979 |
| Relative celerity difference | 3.3% | 6.25% | 3.45% | / |

The longitudinal velocity values obtained from each method are very close. The relative difference which is calculated compared to the velocity obtained by the insertion substitution method is less than 6.5 % of discrepancy as shown in Table.1. These results validate the consistency of our *in-situ* measurement method.

Figure 8 illustrates the evolution of the relative ultrasonic velocity as a function of the relative measured complex viscosity of the silicone oils. It can be noted a linear increase of the ultrasonic velocity with the complex viscosity with a slope around 0.25 ms$^{-2}$Pa$^{-1}$. This result confirms the possibility to monitor the polymer viscosity by longitudinal velocity measurement. As previously mentioned, the 1 Pa.s silicone oil viscosity values show a higher variation.

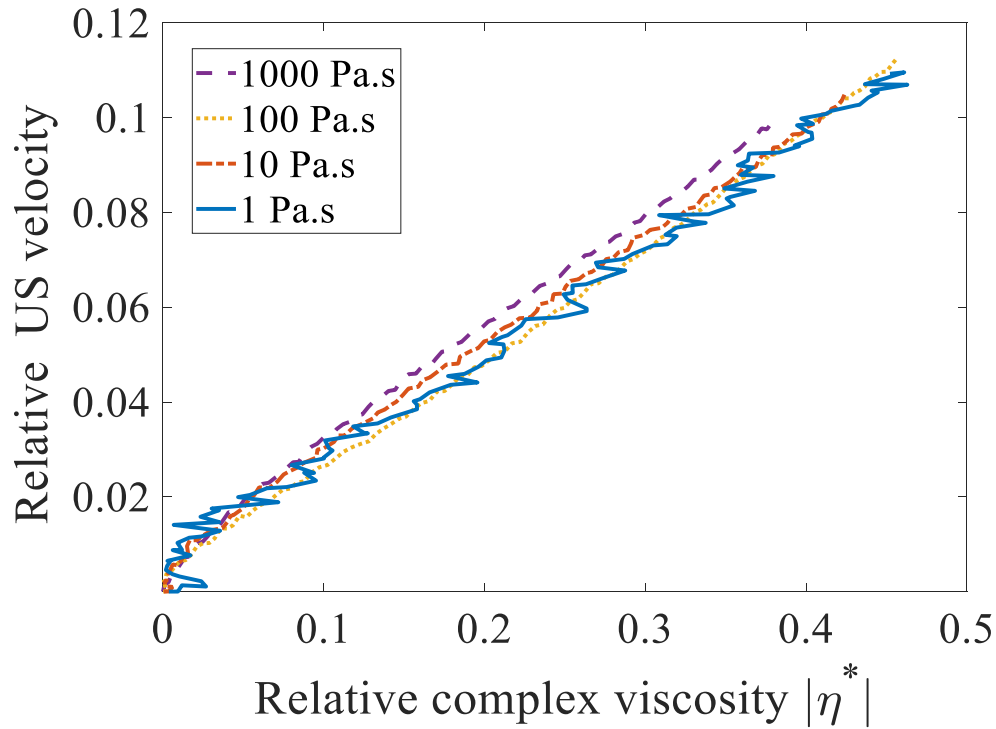

Figure 8. Relative velocity *vs.* relative complex viscosity for **1,10,100 and 1000 Pa.s** silicone oils during a temperature rise from 25°C to 85°.

## 4.2 Results during polyethylene change of state

### 4.2.1. Complex viscosity measurements

Simultaneous rheological measurements of the gap width which is equal to polymer thickness and the complex viscosity $|\eta^*|$ during a heating cycle from 160°C to 200°C and a cooling cycle from 200°C to 115°C are presented in Figures 9(a) and 9(b).

When the polymer is heating, the complex viscosity varies slowly from 6.7 kPa.s to 4.6 kPa.s between 160°C and 200°C. As expected, the viscosity decreases with the temperature increase. Simultaneously, the gap decreases from 1.2mm to 1.15mm.

When the melt polymer is cooling, the complex viscosity slowly increases from 4.6 kPa.s to 10 kPa.s When the temperature reaches 118°C, a fast increase of the complex viscosity is observed. In the same time, the gap width curve versus temperature shows a non-monotonous behavior with an increase of the gap width to 1.24mm at 125°C before to decrease to 1.2mm at 115°C.

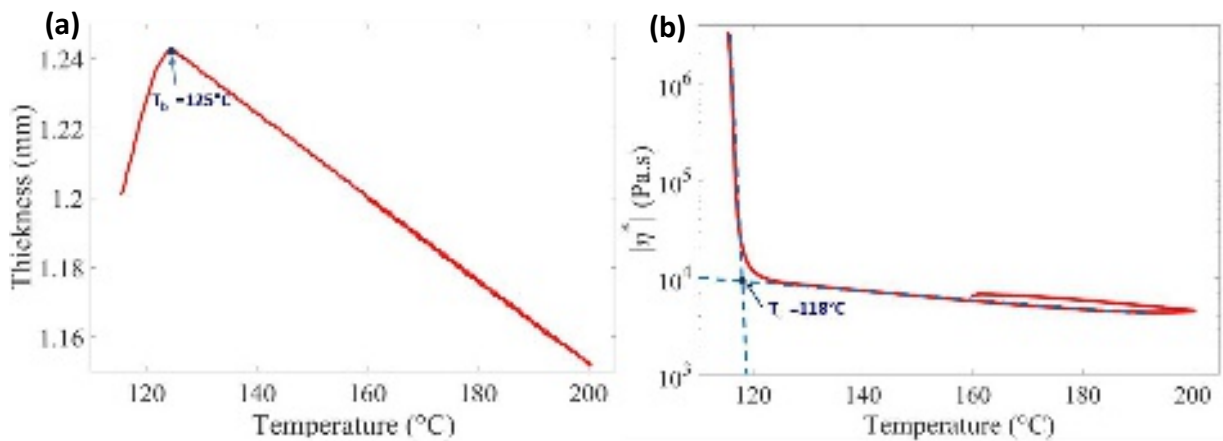

Figure 9. Measurement results of (a) the thickness of MDPE (gap width), $T_b$ is the characteristic point corresponding to the behavior change of the gap width. (b) the complex viscosity ($|\eta^*|$) as a function of temperature during heating (160-200°C) and cooling (200°C- 115°C), $T_c$ is the characteristic point corresponding to the fast increase of the complex viscosity.

### 4.2.2. Simultaneous ultrasonic velocity measurements

Figure 10 represents the cartography of the received signals measured every 6s during 28 minutes corresponding to the cooling of the MDPE from 200°C to 80°C (from top to bottom). The rheological measurements are stopped at $t_m$=18min, which corresponds to a temperature of 115°C because of the MDPE solidification. The ultrasonic measurements are continued until $t_m$=28min (80C°). For better readability, the signals have been fully rectified. When the melt polymer is cooling, the US velocity in the quartz increases which this explains the shift of the TOF that can be seen between 8 and 9µs (the echo reflected from the delay line/quartz interface). From 9.5µs, the different back and forth echoes in the quartz $Q_{q+p,1}$, $Q_{q+p,2}$ and $Q_{q+p,3}$ are observed.
16

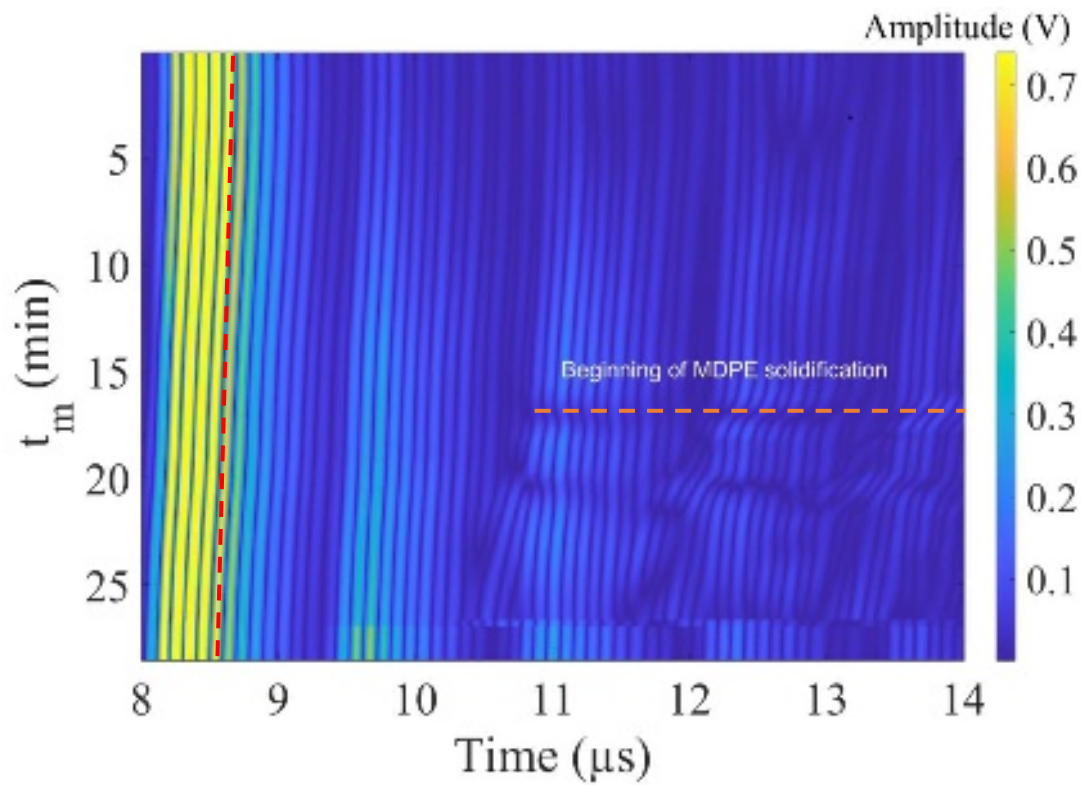

Figure 10. Cartography of received signals as a function of measurement time during MDPE cooling ($t_m$=0min corresponds to 200°C and $t_m$=28mn to 80°C); The beginning of solidification, depicted by the orange dashed line, corresponds to the fast decrease of the TOF of E1(t) at $t_m$=17min and t=11µs).

The echo $E_1(t)$ corresponding to the first back and forth in the MDPE overlaps with the second back and forth in the quartz $Q_{q+p,2}$. The modification of the TOF of E1 is easily observed from $t_m$=17min because of the interferences between this echo and $Q_{q+p,2}$, followed by a sudden decrease in the TOF from 20min. This pattern is repeating at t=12.3µs and at t=13.8µs which corresponds to the back and forth in the MDPE. From $t_m$=27min, the MDPE is solidified and crystallized. It detaches from the quartz plate due to the shrinkage force. The amplitude of the echoes $Q_{q,n}$ clearly increases because of the interface change. This echo $E_1$ is extracted from the signal using the method explained earlier.

Figure 11 shows the results of the ultrasonic velocity measurements as a function of temperature during a heating cycle from 160°C to 200°C and a cooling cycle from 200°C to 115°C. Measurements are achieved simultaneously with the rheological measurements presented in the subsection 4.2.1. When the polymer is heating, a slow decrease from 1380 m/s to 1300 m/s is observed, followed by a slow increase from 1300 m/s at 200°C to 1400 m/s at 130°C. Afterward, sudden increase in US velocity is observed between 130°C and 125°C. From 118°C, the velocity quickly increases until the end of the experiment.

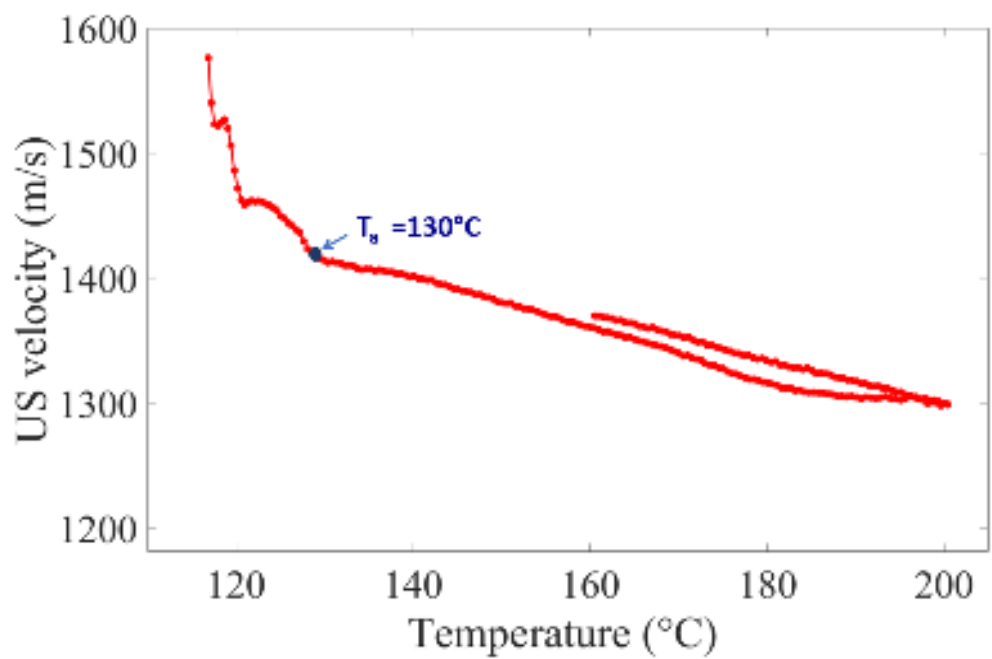

**Figure 11.** Measurement results of the ultrasonic velocity as a function of temperature during heating (160-200°C) and cooling (200°C- 115°C) of the MDPE (section 4.2.1). $T_a$ is the characteristic point corresponding to the slope change of the velocity.

From these measurements, three characteristic points around the solidification onset of the MDPE can be identified. When the melt polymer is cooling, the first point at $T_a$ =130°C (Figure.11) corresponds to the beginning of a fast increase in the ultrasonic velocity. At the second point at $T_b$= 125°C (Figure 9.a) is characterized by a behavior change of the thickness. The last point at $T_c$=118°C (Figure 9.b) corresponds to a fast increase of the complex viscosity. These characteristic points are presents in the ultrasonic velocity curve by a slope change (Figure.11). In fact, three increasing slopes are observed. The first increasing slope from 200°C to 130°C corresponds to the MDPE in a molten state. The second increasing slope faster than the previous from 130°C to 118°C corresponds to the onset state change. The third slope even faster than the previous from 118°C corresponds to the solid state of MDPE. The fast increase of the complex viscosity at the same temperature permits to confirm that the MDPE is in a solid state at 118°C. Ultrasonic velocity appears to be precursor to detect changes in the MDPE state. It allows to detect the solidification onset.

Figure 12 shows the evolution of the US velocity variation as a function of the complex viscosity variation during the heating and the cooling cycles. A linear behavior of the velocity as a function of the complex viscosity in a temperature range from 200°C to 130°C is observed. A correlation between US and rheological measurements can be achieved in this range. After that, both viscosity and velocity quickly increase because of the beginning of the MDPE crystallization

18ACCEPTED MANUSCRIPT
AIP Advances
AIP Publishing
This is the author's peer reviewed, accepted manuscript. However, the online version of record will be different from this version once it has been copyedited and typeset.
PLEASE CITE THIS ARTICLE AS DOI: 10.1063/5.0251850

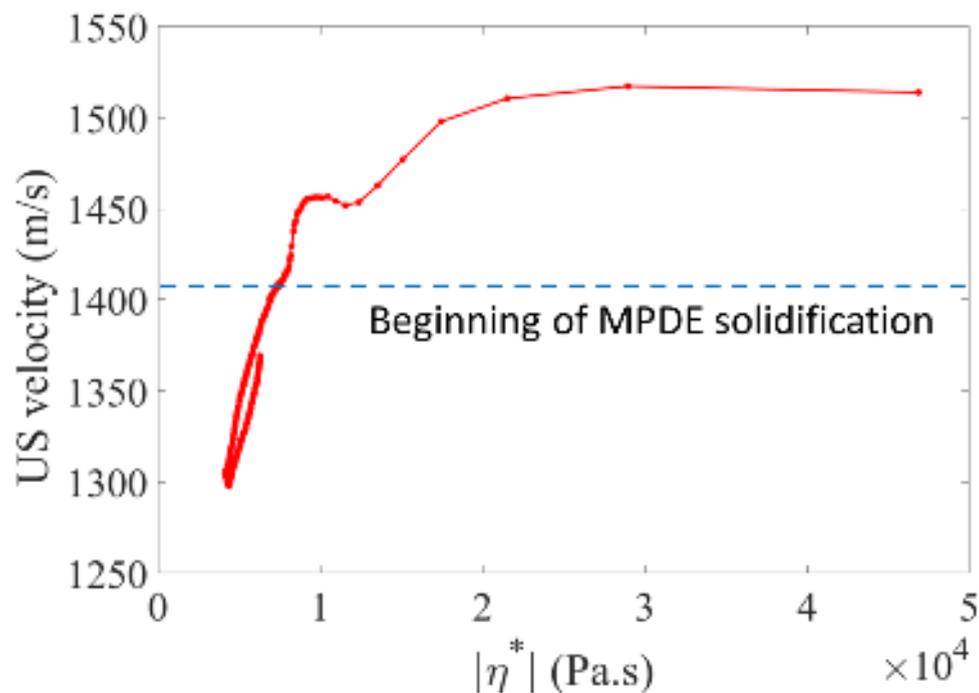

Figure 12. Measurement results of the ultrasonic velocity as a function of the complex viscosity of the MDPE during heating (160-200°C) and cooling (200°C- 115°C) steps. The dashed blue line indicates the beginning of the solidification.

## 5. Conclusion

An ultrasonic instrumentation of an oscillatory rheometer is proposed. It allows a simultaneous measurement of ultrasonic velocity and complex viscosity of the medium density polyethylene at high and varying temperatures up to 200°C. A signal processing method enables the determination of the overlapped polymer echo time of flight by considering of the temperature variation effect. To validate the developed instrumentation and method, the measurements are performed on silicone oils with different calibrated viscosities at 25°C. The comparison with a conventional longitudinal velocity measurement by insertion substitution method demonstrates its robustness. Results of velocity and viscosity measurements on MDPE shows that the ultrasonic parameters appear to be more sensitive to the evolution of the polymer viscosity and to changes in polymer state. It enables an accurate detection of the onset of polymer solidification. Moreover, the ultrasonic velocity allows to monitor the evolution of polymer viscosity with temperature. The potential of ultrasound in understanding and controlling the mechanisms involved in the crystallization of semi-crystalline polymers could be further investigated. The study paves the way of ultrasonic real-time monitoring of the rotational molding process.

## Acknowledgements
Rotational molding process is employed for manufacturing polyethylene liner of high-pressure hydrogen storage tanks, storage technology could facilitate the development of clean heavy hydrogen mobility. The authors acknowledge the ANR – FRANCE (French National Research Agency) for its financial support of the Project ULHYS **ANR-21-CE05-0019-01** "Simulation of rotomolded polymer viscosity for hydrogen by ultrasonic testing method".